\documentclass[5p,twocolumn]{elsarticle}
\usepackage{graphicx,latexsym}
\usepackage{amssymb,amsmath,bm}
\usepackage{subfigure}
\usepackage{braket}

\usepackage{hyperref}
\hypersetup{
    pdfnewwindow=true,       
    colorlinks=true,         
    linkcolor=blue,          
    citecolor=blue,          
    filecolor=magenta,       
    urlcolor=black           
}

\def\eq#1{Eq.\ (\ref{#1})}
\def\mb#1{\mbox{\boldmath$#1$}}
\def\fig#1{Fig.\ \ref{#1}}

\journal{}

\begin{document}

\begin{frontmatter}

\title{Single-photon controlled thermospin transport in a resonant ring-cavity system}

\author[a1,a2,a3]{Nzar Rauf Abdullah}
\ead{nzar.r.abdullah@gmail.com}
\address[a1]{Physics Department, College of Science, University of Sulaimani, Kurdistan Region, Iraq}
\address[a2]{Komar Research Center, Komar University of Science and Technology, Sulaimani City, Iraq}
\address[a3]{Science Institute, University of Iceland, Dunhaga 3,
        IS-107 Reykjavik, Iceland}

\author[a3]{Vidar Gudmundsson}
\ead{vidar@hi.is}

%

\begin{abstract}

Cavity-coupled nanoelectric devices hold great promise for quantum technology based on 
coupling between electron-spins and photons. 
In this study, we approach the description of these effects through the modeling of a 
nanodevice using a quantum master equation.
We assume a quantum ring is coupled to two external leads with different temperatures 
and embedded in a cavity with a single photon mode. Thermospin transport of the ring-cavity system 
is investigated by tuning the Rashba coupling constant and the electron-photon coupling strength.
In the absence of the cavity, the temperature gradient of the leads causes 
a generation of a thermospin transport in the ring system. 
It is observed that the induced spin polarization has a maximum value 
at the critical value of the Rashba coupling constant 
corresponding to the Aharonov-Casher destructive interference, 
where the thermospin current is efficiently suppressed.
Embedded in a photon cavity with the photon energy close to a resonance with the energy spacing between 
lowest states of the quantum ring, a Rabi splitting in the energy spectrum is observed. Furthermore, 
photon replica states are formed leading to a reduction in the thermospin current. 
 
\end{abstract}


\end{frontmatter}

\section{Introduction}
Thermoeletric properties have been so far investigated mainly in nanoscale 
systems to achieve high thermoelectric efficiency that would be useful 
for energy harvesting~\cite{Weihua_2016,WEI_2018}. 
To obtain a high thermoelectric efficiency, thermoelectrically active materials are used.
These materials should have high electrical conductivity and low thermal conductivity. 
High electrical conductivity can be obtained by increasing the carrier mobility or 
their concentration that can be influenced in quantum structures.
Consequently, the figure of merit and Seebeck effect can be enhanced 
in nanodevices~\cite{Zhifeng_npj_2016,Heremans25072008,Hochbaum455.778}.
Generally, there is a challenge in the conventional thermoelectric material because if the electrical conductivity is enhanced 
the thermal conductivity is increased as well. As a result, the Seebeck effect and the device efficiency are decreased.

An advantageous method relying on the Spin Seebeck effect has been used to decouple electrical conductivity 
from thermal conductivity. So the electrical and thermal conductivity 
can be separately controlled concurrently~\cite{Maekawa_2008,Xiong_2015}. 
In 2008, Saitoh and \textit{et.\ al.}\ discovered 
the spin Seebeck effect when heat is applied to a magnetized metal. 
In a magnetically active material electrons reconfigure themselves according to their spin. 
In this way, unlike in a conventional electron transport, this rearrangement does not
create heat as a waste product. 
The spin Seebeck effect can lead the way to the growth of smaller, faster and more energy-efficient 
microchips as well as spintronics devices~\cite{Maekawa_2013_SpinSeebeckEffect,Uchida2016,Liu2011}.

On the other hand, the influences of light on thermoelectric effects have been 
investigated and shown that the thermoelectric power can be enhanced 
by increasing the intensity of light~\cite{Mitra_2008}.
It was also found that a polarized light can induce a Fano-like resonance 
in the thermal conductance~\cite{BAI2014190}. 
Therefore, The thermopower and the figure of merit may be enhanced 
near a Fano-like resonance. In addition, the polarized light and the increase of magnetic 
polarization may lead to a better thermoelectric performance,
especially, a significant increase of the spin thermal efficiency may be obtained.

Influenced by the aforementioned studies, we try to explore the influences of a quantized photon field 
on thermospin transport through a quantum ring including the Rashba spin-orbit coupling. 
We model a quantum ring system coupled to two leads with different temperatures. 
The ring system is embedded in a cavity with a linearly polarized photon field. 
In our previous publications, we have seen that both thermoelectric and heat currents 
can be controlled by a polarized photon field~\cite{Nzar_ACS2016,Abdullah2017,ABDULLAH2018199,ABDULLAH2018}. 
The aim of our study here is twofold. First, we induce a thermospin current through a multi-level 
quantum ring and see the influences of the Rashba spin-orbit coupling on thermospin current using 
a quantum master equation.
Second, we show how the spin-polarization and thermospin current can 
be controlled by a single photon in the cavity.

The paper is organized as follows: In Sec.~\ref{Sec:II}, we present
the model describing a quantum ring coupled to a photon cavity. 
Section \ref{Sec:III} shows the numerical results and
discussion. Concluding remarks are addressed in Sec.~\ref{Sec:IV}.

\section{Model and Theory}\label{Sec:II}

In this section, we first present the Hamiltonian of the system, 
and subsequently apply the general master equation formalism to calculate 
the thermospin-polarized current. 

\subsection{Hamiltonian of the system}

The Hamiltonian of a quantum ring system coupled to a cavity can be expressed as

\begin{equation}
 H = H_e  + H_{e-\gamma} + H_{\rm \gamma},
\label{Eq_1}
\end{equation}
where the Hamiltonian of the electronic part $\hat{H}_{e}$ and the electron-photon interaction $H_{e-\gamma}$ together can be defined as
\begin{eqnarray}
 \hat{H}_{e} + H_{e-\gamma} &=&\int d^2 r\; \hat{\mathbf{\Psi}}^{\dagger}(\mathbf{r})\left[\left(\frac{\hat{\mathbf{p}}^2}{2m^{*}} +V_r(\mathbf{r})\right) + H_{Z} \right. \nonumber \\
 &+&\left. \hat{H}_{R}(\mathbf{r})\right]\hat{\mathbf{\Psi}}(\mathbf{r})
 +\hat{H}_{ee},
 \label{Eq_2}
\end{eqnarray}
and the Hamiltonian of the free photon field in the cavity is 
\begin{equation}
 H_{\rm \gamma} = \hbar \omega_{\gamma} \hat{a}^{\dagger}\hat{a}.
\end{equation}

Herein, $\hat{\mathbf{\Psi}}(\mathbf{r})$ is the spinor vector~\cite{ABDULLAH2018199}, and $\hat{\mathbf{p}}$ is
the momentum operator of the quantum ring system coupled to the photon cavity which can be written as
\begin{equation}
 \hat{\mathbf{p}}(\mathbf{r}) = \frac{\hbar}{i}\nabla +\frac{e}{c} \left[\hat{\mathbf{A}}(\mathbf{r}) +
 \hat{\mathbf{A}}_{\gamma}(\mathbf{r})\right],   
 \label{Momentum_P}
\end{equation}
where the vector potential of the external perpendicular magnetic field is $\hat{\mathbf{A}}(\mathbf{r}) = -By\hat{x}$  with 
$\mathbf{B} = B \hat{\mb{z}}$, and the vector potential of the photons in the cavity is $\hat{\mathbf{A}}_{\gamma}(\mathbf{r})$ 
which can be introduced in terms of the photon creation ($\hat{a}^{\dagger}$) and annihilation ($\hat{a}$) operators as
\begin{equation}
 \hat{\mathbf{A}}_{\gamma}=A(\mathbf{e}\hat{a}+\mathbf{e}^{*}\hat{a}^{\dagger}) ,
 \label{vec_pot}
\end{equation}
with $\mathbf{e}= \mathbf{e}_x$ for the $x$-polarized and $\mathbf{e}= \mathbf{e}_y$ for the  
$y$-polarized photon field~\cite{Nzar_IEEE_2016}. The potential that forms the quantum ring is $V_r(\mathbf{r})$
and $H_{Z}$ is the Zeeman Hamiltonian~\cite{ABDULLAH2018199}.
The strength of the vector potential of the photons $A$ is defined by the electron-photon 
coupling constant $g_{\gamma} = eA\Omega_w a_w/c$, where 
$\Omega_w = (\omega^2_c + \Omega^2_0)^{\frac{1}{2}}$ is the effective characteristic frequency 
with $\Omega_0$ the frequency of the confined electron in the $y$-direction, $\omega_c$ is 
the cyclotron frequency and $a_w$ the effective magnetic length.

Furthermore, $\hat{H}_{R}(\mathbf{r})$ is the Rashba-spin orbit interaction
\begin{equation}
 \hat{H}_{R}(\mathbf{r})=\frac{\alpha}{\hbar}\left( \sigma_{x} \hat{p}_y(\mathbf{r}) -\sigma_{y} \hat{p}_x(\mathbf{r}) \right) ,
 \label{H_R}
\end{equation}
with $\alpha$ the Rashba spin-orbit (RSO) coupling constant that can be tuned by an external electric field, 
and $\sigma_{x}$ and $\sigma_{y}$ are the Pauli matrices.
The last term of \eq{Eq_2} is $\hat{H}_{ee}$ which accounts for the electron-electron interaction of the 
quantum ring system~\cite{Nzar_IEEE_2016,nzar27.015301}.
The Hamiltonian presented in \eq{Eq_1} is used to obtain the energy spectrum of the quantum ring-cavity system 
using a numerically exact diagonalization technique \cite{PhysRevB.82.195325,0953-8984-30-14-145303}.

\subsection{Transport Formalism}

To describe the transient electron transport through the quantum ring system, we use 
a time-convolution-less generalized master equation (TCL-GME)~\cite{Arnold13:035314,Arnold2014,Thorsten2014}.
The TCL-GME is local in time and satisfies the positivity for the many-body state occupation 
described by the reduced density operator (RDO). 
The RDO of the system quantum ring system $\hat{\rho}_S$, in terms of the total density matrix $\hat{\rho}_T$, 
is defined as 
\begin{equation} 
 \hat{\rho}_S(t) = {\rm Tr}_l \big[\hat{\rho}_T(t) \big] ,
 \label{RDO}
\end{equation}
where $l \in \{L,R\}$ indicates the two electron reservoirs, 
the left (L) and the right (R) leads, respectively.

In our study, we integrate the GME to the point in time $t = 220$~ps, late in the transient regime when the total system 
is approaching the steady state.
 
The RDO is utilized to calculate the spin-polarization and thermospin current. 
We define the spin polarization $S_i$ of the quantum ring system
in $i=x,y,z$ direction. Thus, the spin polarization operator is
\begin{equation}
 \hat{S}_i = \int d^2r \hat{n}^i(\mathbf{r}) ,
\end{equation}
with $\hat{n}^i(\mathbf{r})$ the spin polarization density operator for the spin polarization 
$S_i$~\cite{ARNOLD2014170}. 
In addition, the top local thermospin current 
($I^{{\rm th},i}_{\rm t}$) through the upper arm ($y>0$) of the quantum ring system can be introduced as
\begin{equation}
 I^{{\rm th},i}_{\rm t}(t) = \int_{0}^{\infty} dy \, j_x^{{\rm th},i}(x=0,y,t) 
 \label{eq:topcurrent}
\end{equation}
and the bottom local thermospin polarization current ($I^{{\rm th},i}_{\rm b}$) 
through the lower arm ($y<0$) of the quantum ring system
\begin{equation}
 I^{{\rm th},i}_{\rm b}(t) = \int_{-\infty}^{0} dy \, j_x^{{\rm th},i}(x=0,y,t) ,
  \label{eq:bottomcurrent}
\end{equation}
where the spin polarization current density 
is 
\begin{equation}
 \mathbf{j}^{th,i}(\mathbf{r},t) = {{j^{th,i}_x(\mathbf{r},t)}\choose{j^{th,i}_y(\mathbf{r},t)}} 
 = {\rm Tr} \Big[\hat{\rho}_s(t) \hat{\mathbf{j}}^{th,i}(\mathbf{r}) \Big]
\end{equation}
calculated by the expectation value of the spin polarization 
current density operator~\cite{ARNOLD2014170}.

Finally, the total local (TL) thermospin polarization current is obtained from 
the top and the bottom thermospin current polarization
\begin{equation}
 I_{tl}^{{\rm th},i}(t) =  I^{{\rm th},i}_{\rm t}(t) + I^{{\rm th},i}_{\rm b}(t) .
\end{equation}

The TL-thermospin current is 
related to non-vanishing spin-polarization sources, 
and the circular local (CL) thermospin polarization current 
is 
\begin{equation}
 I_{cl}^{{\rm th},i}(t) = \frac{1}{2} \Big[ I^{{\rm th},i}_{\rm b}(t) - I^{{\rm th},i}_{\rm t}(t)\Big].
\end{equation}

In the result section, we present the main results on the thermospin transport 
in the quantum ring system and the influence of the photon field on the quantum ring system.

\section{Results}\label{Sec:III}

In this section the numerical results are shown for a ring-cavity system
including the Rashba spin-orbit interaction and the electron-photon interaction. 
The quantum ring and the leads 
are made of a GaAs-based material with electron effective mass  $m^* = 0.067 m_e$ and 
the relative dielectric material $\kappa = 12.4$.

An external perpendicular 
magnetic field with strength $B = 10^{-5}$~T is applied to the 
total system including the leads. 
We assume a very weak external magnetic field 
to avoid spin degeneracy, and have it weak enough to prevent 
creating a circular motion due to a Lorentz force in the quantum
ring system.  The main goal of the study here is to show the ``real''
circular local (CL) thermospin current due to the Rashba effect in the
quantum ring system.  Therefore we choose the low strength of the external magnetic
field and tune the Rashba coupling constant.
This assumed external magnetic field 
is out of the Aharonov-Bohm (AB) regime because 
the area of the ring structure is $A = \pi a^2 \approx 2 \times 10^4$ nm$^2$ leading to 
a magnetic field $B_0 = \phi_0/A \approx 0.2$~T corresponding to one flux
quantum $\phi_0 = hc/e$~\cite{Arnold2014,Nzar_2016_JPCM}. 

The ring system is parabolically confined in the $y$-direction with characteristic energy 
$\hbar \Omega_0 = 1.0$~meV. 
It gives a broad quantum ring. However, the broad ring geometry together
with the spin degree of freedom, and the spin-orbit interactions require
a substantial computational effort. Transport properties linked to
spin-orbit interactions can be clearly realized in a broad quantum ring.

It is assumed that the cavity initially contains one photon
with linear polarization.
It is also considered that the left and the right leads have the same chemical potential ($\mu_L = \mu_R = \mu$),
but are at different temperatures, which induce a thermal transport in the quantum ring system. 

\subsection{Energy spectrum}

\begin{figure}[htb]
\centering
    \includegraphics[width=0.4\textwidth,angle=0]{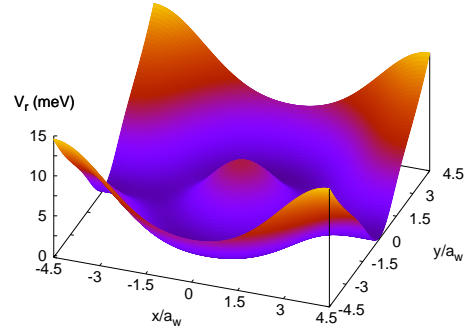}
 \caption{(Color online) The potential $V_r(\mathbf{r})$ defining the central ring
       system that will be coupled diametrically to the semi-infinite left and right leads
       in the $x$-direction.}
\label{fig01}
\end{figure}

The potential defining the quantum ring system and its energy spectrum are demonstrated here.
Figure \ref{fig01} shows the potential of the ring where the top arm 
is located in the positive $y$-axis and the bottom arm is in the negative $y$-axis. 
It should be mentioned that the electrons are mainly transferred through the ring system in the $x$-direction
by the thermal bias.

\begin{figure}[htb]
\centering
    \includegraphics[width=0.3\textwidth,angle=0]{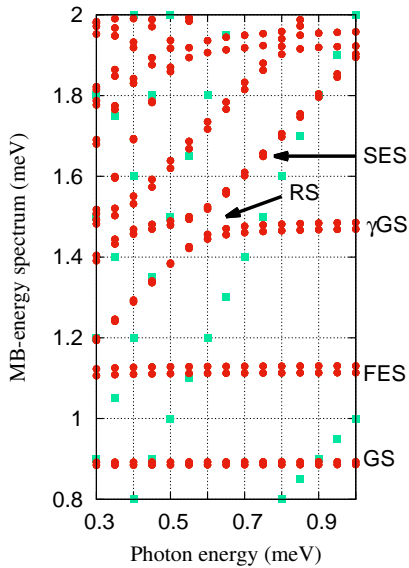}
 \caption{(Color online) Many-Body (MB) energy spectrum of the ring system versus 
          the photon energy $\hbar \omega_{\gamma}$.
          The green rectangles are the zero-electron states (0ES),
          the red circles indicate the one-electron states (1ES), 
          and the two electron states are located at the upper part of the energy spectrum (now shown).
          The black arrow indicates the Rabi-splitting (R-S) between the second-excited state and 
          the one-photon replica of the ground state. 
          The Rashba coupling constant is fixed at $\alpha = 14.0$~meV, the electron-photon coupling strength
          is $g_{\gamma} = 0.05$~meV, and the photon is linearly polarized in the $x$-direction. 
          The magnetic field is $B = 10^{-5}$~T, and $\hbar \Omega_0 = 1.0~{\rm meV}$.}
\label{fig02}
\end{figure}

The energy spectrum of the quantum ring system as a function of the photon energy is plotted in 
\fig{fig02} where the electron-photon coupling strength is $g_{\gamma} = 0.05$~meV and the photon is linearly polarized in 
the $x$-direction. The Rashba coupling constant is fixed at $\alpha = 14.0$~meV nm, which is a critical value of the 
Rashba coupling constant described later.
The energy states around $0.885$~meV and $1.112$~meV are the one-electron ground state (GS) and the first-excited state (FES), 
respectively. The aforementioned states are double states due to the spin-orbit interaction including both spin-up and spin-down states.
The energy of the second-excited state (SES) is monotonically increased with increasing photon energy.
The energy value of the SES is $1.4$~meV at photon energy $\hbar \omega_{\gamma} = 0.3$~meV and it is enhanced to $1.9$~meV 
at $\hbar \omega_{\gamma} = 1.0$~meV. 
In addition, the SES is in resonant with the first photon replica of the ground state ($\gamma$GS).
The energy value of $\gamma$GS is $1.2$~meV at 
$\hbar \omega_{\gamma} = 0.3$~meV and it becomes $1.49$~meV at $\hbar \omega_{\gamma} = 1.0$~meV.
The resonant states, the SES and the $\gamma$GS, form a Rabi-splitting (RS) in the energy spectrum.
The strongest Rabi-effect is recorded at the photon energy $\approx 0.55$~meV (black arrows). 

\begin{figure}[htb]
\centering
    \includegraphics[width=0.23\textwidth,angle=0]{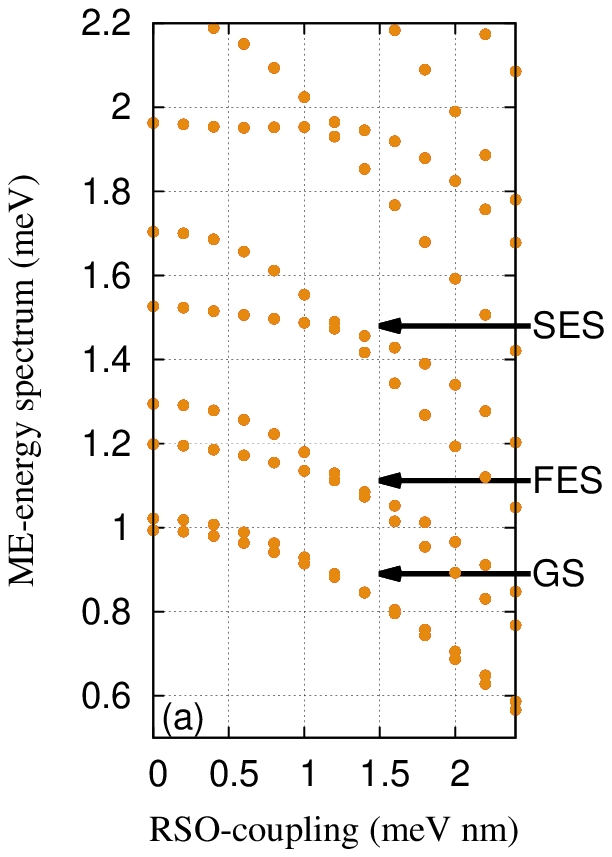}
    \includegraphics[width=0.23\textwidth,angle=0]{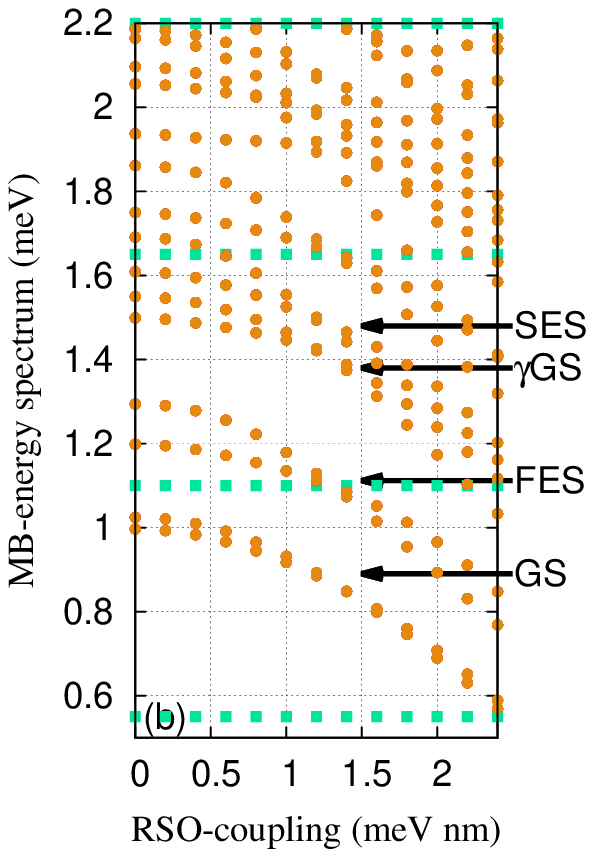}
 \caption{(Color online) Many-Electron (ME) energy spectrum (a) and Many-Body (MB) energy spectrum (b)
          of the ring system versus the Rashba spin-orbit (RSO) coupling constant 
          ($\alpha$).
          The green rectangles are the zero-electron states (0ES),
          the brown circles indicate the one-electron states (1ES), 
          and the two electron states are located at the upper part of the energy spectrum (now shown).
          The photon energy is $\hbar \omega_{\gamma} = 0.55$~meV, the electron-photon coupling strength
          is $g_{\gamma} = 0.05$~meV, and the photon in the cavity is linearly polarized in the $x$-direction. 
          The magnetic field is $B = 10^{-5}$~T, and $\hbar \Omega_0 = 1.0~{\rm meV}$.}
\label{fig03}
\end{figure}

We study the thermospin transport in the strong Rabi effect regime.
Therefore, we fix the photon energy at $\hbar \omega_{\gamma} = 0.55$~meV in our calculations from now on, 
and investigate the properties of the thermospin transport of the system.
The many-electron (ME) energy (a) and the Many-Body (MB) energy (b) for the photon dressed electron states 
of the quantum ring system versus 
the RSO-coupling are shown in \fig{fig03}. 
In the absence of the cavity (\fig{fig03}(a)), 
the energy of the one-electron states decreases with increasing RSO-coupling and 
crossings of the one-electron states are formed at $\alpha \approx [10-15]$~meV nm. 
The crossing of the states corresponds to the Aharonov-Casher (AC) destructive phase interference 
in the quantum ring system. 
In our study, we focus on the three lowest degenerate energy states which are 
the the components of the GS at $E_{\rm GS} \backsimeq 0.88$~meV, 
FES at $E_{\rm FES} \backsimeq 1.112$~meV, and 
SES at $E_{\rm SES} \backsimeq 1.47$~meV, respectively.

The MB-energy spectrum of the quantum ring system in the presence of the cavity is shown in \ref{fig03}(b)
where the photon energy is $\hbar \omega_{\gamma} = 0.55$~meV
and electron-photon coupling strength $g_{\gamma} = 0.05$~meV. 
Comparing to the ME-energy shown in \fig{fig03}(a),
in addition to the degenerate states, the photon replica states are formed.
The energy spacing between the photon replicas is approximately equal to the 
photon energy at low electron-photon coupling strength $g_{\gamma} = 0.05$~meV. 
For instance, the $\gamma$GS is formed near the 
SES and the energy spacing between GS and SES is 
approximately equal to the photon energy $E_{\rm SES} - E_{\rm GS} \approx \hbar \omega_{\gamma} = 0.55$~ meV.
Under this condition, the quantum ring system is resonant with the photon field~\cite{ABDULLAH2018199}.
In addition, we should mention that the first subband energy of the semi-infinite leads is located at $1.0$~meV (not shown).
Therefore, the GS energy of the quantum ring system does not play an important role in the transport. 

In addition, the range of RSO is $\alpha = [0:24]$ meV nm which is a reasonable 
and applicable range for GaAs materials. 
The parameter $\alpha$ depends on the electric field. An
electric field can be generated in a heterostructure with two layers
either by the intrinsic potential at the interface or as an external
field. The Rashba parameter depends on that electric field,
and thus it can be varied in the selected range here~\cite{ARNOLD2014170,Tuan2014}.

\subsection{Transport properties in the absence of the cavity}
\begin{figure}[htb]
\centering
    \includegraphics[width=0.45\textwidth,angle=0,bb=57 85 410 210]{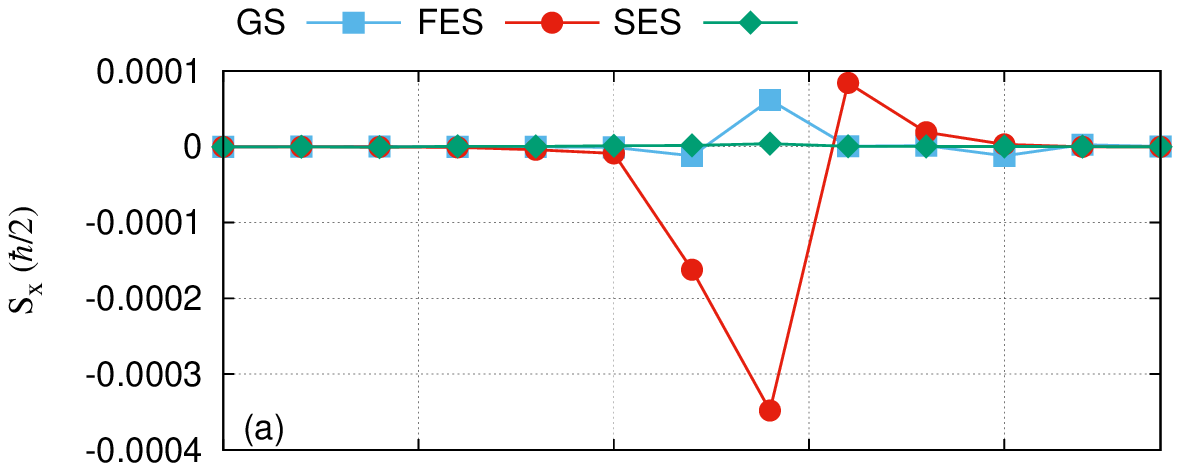}\\
    \includegraphics[width=0.45\textwidth,angle=0,bb=50 85 410 208]{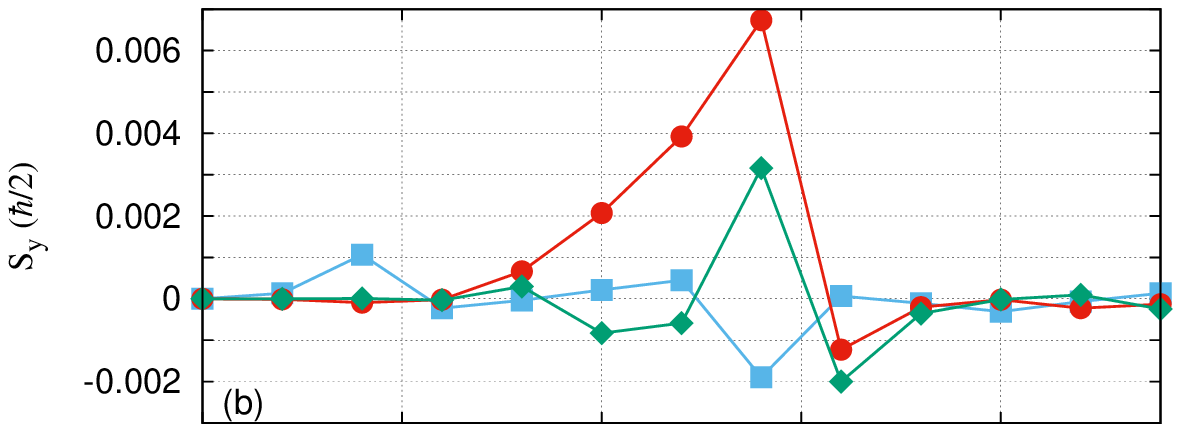}\\
    \includegraphics[width=0.45\textwidth,angle=0,bb=64 60 408 207]{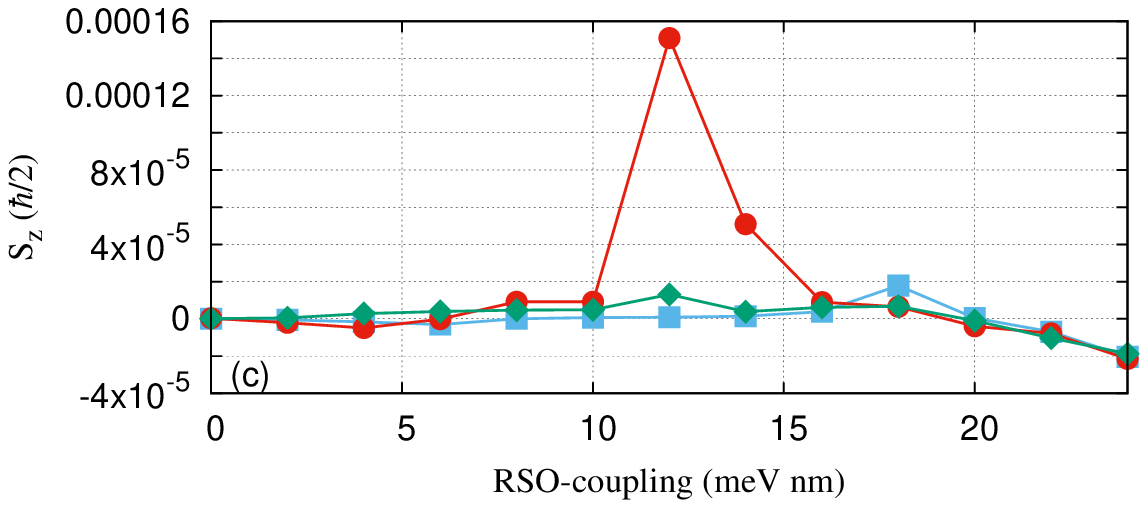}
 \caption{(Color online) Spin polarization $S_x$ (a), $S_y$ (b) and $S_z$ (c) of the quantum ring 
  system without the photon field versus the RSO-coupling constant. 
  The temperature of the left (right) lead is fixed at $T_{\rm L} = 0.41$~K ($T_{\rm R} = 0.01$~K) implying a
  thermal energy $k_B T_{\rm L} = 0.35$~meV ($k_B T_{\rm R} = 0.00086$~meV), respectively.       
  The magnetic field is $B = 10^{-5}$~T, and $\hbar \Omega_0 = 1.0~{\rm meV}$.}
\label{fig04}
\end{figure}

In this section we investigate the transport properties of the quantum ring system without the photon field.
To calculate the spin-polarization and the thermospin current for the three lowest energy states,  
the chemical potential of the leads are fixed at $\mu_L = \mu_R = 0.91$, $1.112$ and $1.47$~meV 
for the GS, FES and SES calculations, respectively.
The spin polarization, $S_x$ (a), $S_y$ (b) and $S_z$ (c), of the electrons versus the RSO-coupling constant is shown 
in \fig{fig04} for the three lowest energy states of the quantum ring system without the photon field.
The non-vanishing spin-polarization in the range of $\alpha = [10-15]$~meV nm 
corresponds to the location of degenerate energy states (crossing energy states) shown in \fig{fig03}(a) 
and the destructive AC interference.
Furthermore, it appears that the spin-polarization in both $x$- and $z$-directions is much smaller 
than that of $y$-direction ($\sim 10$ times smaller) in the selected range of the Rashba coupling constant
$[10-15]$~meV nm. The reason is that the main transport and canonical momentum are along the $x$-direction, 
and thus the effective magnetic field of Rashba effect should be parallel to the $y$-direction. 
As a result, a higher spin-polarization in the $y$-direction ($S_y$) is induced in the system~\cite{Arnold2014,ARNOLD2014170}.  

We note that the spin-polarization of the GS (blue rectangles) are very small comparing to the FES and SES 
which is due to the position of the GS located below the first subband of the leads.
In addition, the spin-polarization of the FES is higher than that of the SES.

\begin{figure}[htb]
\centering
    \includegraphics[width=0.45\textwidth,angle=0,bb=50 95 410 210]{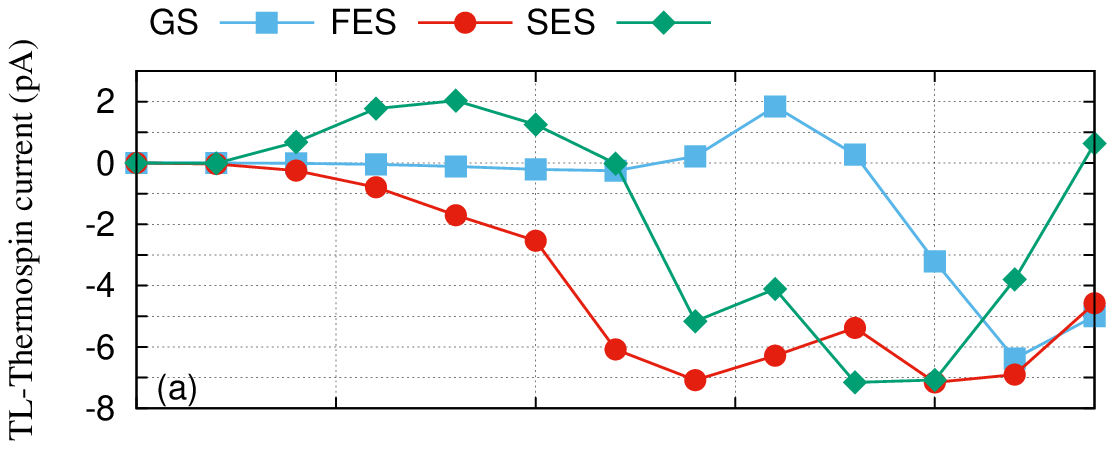}\\
    \includegraphics[width=0.45\textwidth,angle=0,bb=65 55 410 204]{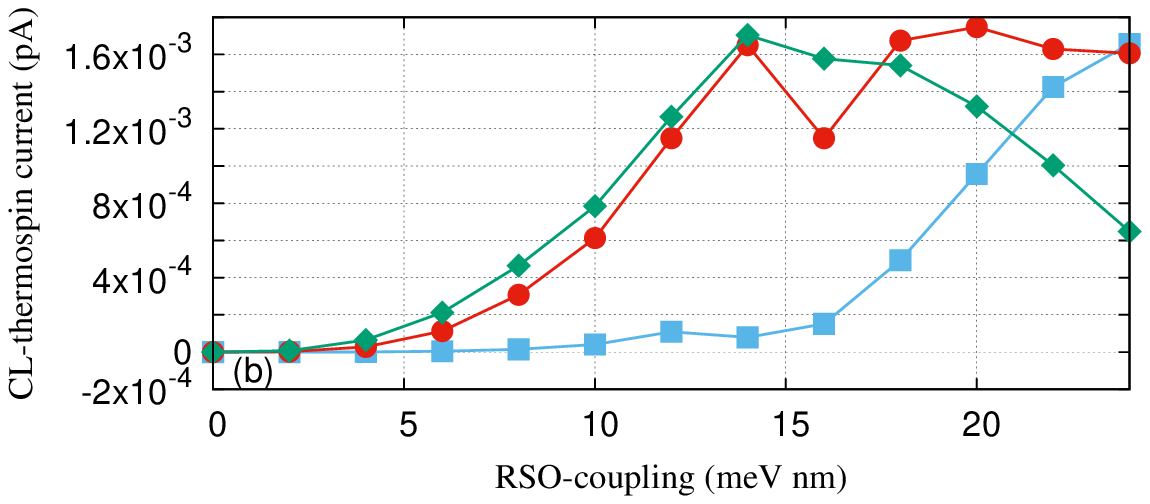}\\
 \caption{(Color online) (a) TL-Thermospin
polarization current, and (b) CL-Thermospin polarization
current in the $y$-direction versus the RSO-coupling for three lowest energy states, GS (blue rectangles),
FES (red circles) and SES (green diamonds) in the absence
of the photon field. The temperature of the left (right) lead is
fixed at $T_L = 0.41$~K ($T_R = 0.01$ K) implying thermal energy
$k_B T_L = 0.35$~meV ($k_B T_R = 0.00086$~meV), respectively. The
magnetic field is $B = 10^{−}5$~T, and $\hbar \Omega_0 = 1.0$ meV.}
\label{fig05}
\end{figure}

Since the $S_y$ is dominant in the quantum ring system comparing to both the 
$S_x$ and $S_z$, we only focus on the properties of the thermospin current in the $y$-direction. 
The TL-Thermospin current (a) and CL-Thermospin current (b) 
as a function of the RSO-coupling are presented in \fig{fig05} 
for the three lowest states of the quantum ring without the cavity.
It is clearly seen that the GS thermospin currents 
up to $\alpha \simeq 15.0$~meV nm are almost zero due to the position of the GS 
located below the first subband of the leads.
For the same range of $\alpha$, the FES and SES are further active in the transport and we notice that
the TL-Thermospin current of the FES and SES has a pronounced minimum value and the CL-Thermospin current has a maximum value at 
the RSO-coupling $\alpha \simeq 14.0$~meV nm corresponding to a destructive AC interference.

We should mention that the CL-Thermospin currents after $\alpha = 15$~meV nm 
is slowly increasing due to the contribution of the FES to the transport at higher RSO-coupling ($\alpha > 15$).
The aforementioned explanations indicates that both FES and SES are active in the transport but with different weights.

\subsection{Transport properties in the presence of the cavity}

\begin{figure}[htb]
\centering
    \includegraphics[width=0.45\textwidth,angle=0,bb=62 50 410 230]{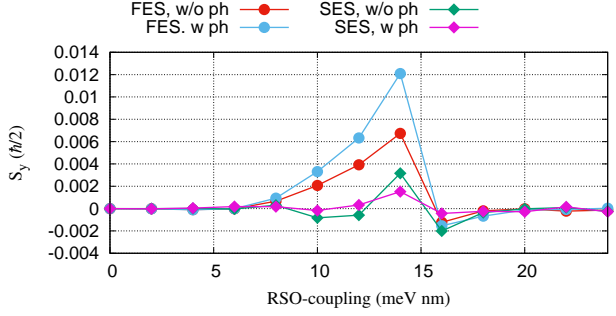}
 \caption{(Color online) Shows the spin polarization $S_y$ versus the RSO-coupling of the quantum ring
  system without photon field (w/o ph) for the FES (red circles) and the SES (green diamonds),
  and with photon field (w ph) for the FES (blue circles) and SES (magenta diamonds).
  The photon energy is $\hbar \omega_{\rm \gamma} = 0.55$~meV and the electron-photon coupling strength is $g_{\rm \gamma} = 0.05$~meV.
  The temperature of the left (right) lead is fixed at $T_{\rm L} = 0.41$~K ($T_{\rm R} = 0.01$~K) implying a
  thermal energy $k_B T_{\rm L} = 0.35$~meV ($k_B T_{\rm R} = 0.00086$~meV), respectively.       
  The magnetic field is $B = 10^{-5}$~T, and $\hbar \Omega_0 = 1.0~{\rm meV}$.}
\label{fig06}
\end{figure}

We now assume the quantum ring system is coupled to the photon field. The energy spectrum of the ring-cavity system 
was shown in \fig{fig03}(b) where the photon field is linearly polarized in 
the $x$-direction. The cavity forms photon replica states influencing the thermospin transport properties. 
Since the FES and SES are the most active state in transport, we focus only on the transport properties of these two states here.
We fix the chemical potential of the leads at $\mu_L = \mu_R = 1.112$~meV for the FES and 
$1.47$~meV for the SES calculations.
The photon energy is fixed at $0.55$~meV which is approximately equal to the energy spacing between the 
GS and the SES. We can thus say that the SES of the quantum ring system is resonant with the photon cavity while 
the FES is off-resonant.
Figure \ref{fig06} shows the $S_y$ spin polarization of the quantum ring system without (w/o ph) and with (w ph) 
photon field. 
We have mentioned, that in the absence of the photon,
the Rashba effective magnetic field should be parallel to the $y$-direction and induce
a spin polarization in the $y$-direction. 
In the presence of the photon field, a kinetic momentum in the $x$-direction is added to the electrons
(see \eq{Momentum_P}), therefore, the $S_y$ spin polarization should increase 
with the $x$-polarized photon field. This can be clearly seen in the $S_y$ spin polarization of the FES, off-resonant regime.
But in the on-resonant regime, the $S_y$ spin polarization of the SES is
slightly decreased in the presence of the photon field which is a direct consequence of the Rabi effect.
Similar effect has been observed for the thermoelectric current in a quantum wire~\cite{Nzar_ACS2016}.

\begin{figure}[htb]
\centering
    \includegraphics[width=0.45\textwidth,angle=0,bb=50 95 410 210]{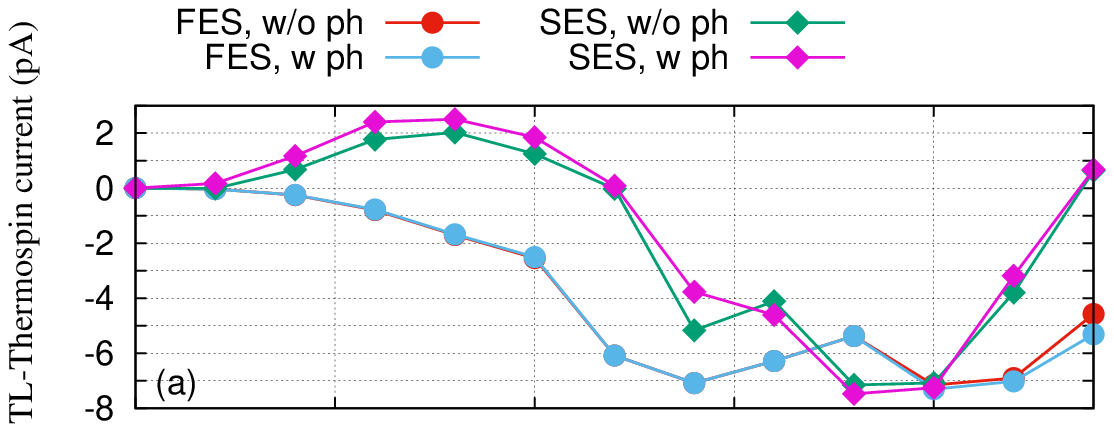}\\
    \includegraphics[width=0.45\textwidth,angle=0,bb=65 55 408 204]{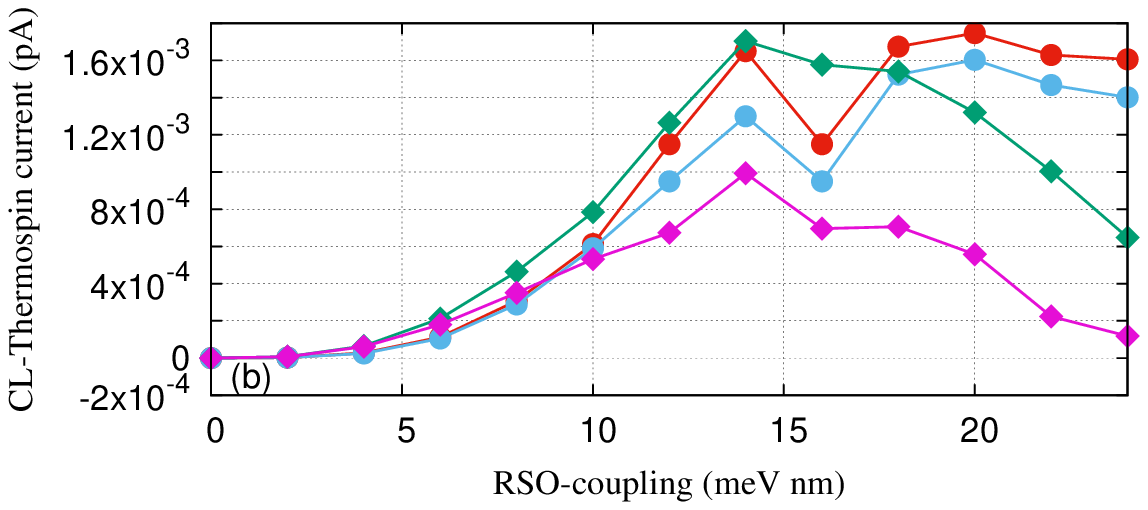}\\
 \caption{(Color online) (a) TL-Thermospin current and (b) CL-Thermospin current versus the RSO-coupling are plotted 
  for the SES of the quantum ring system without photon field (w/o ph) (green circles) and with photon field (w ph) 
  for three different values of the electron-photon coupling strength $g_{\gamma} = 0.05$ (magenta diamonds), 
  $0.1$ (red rectangles) and $0.15$~meV (blue triangles)
  The photon energy is $\hbar \omega_{\rm \gamma} = 0.55$~meV.
  The temperature of the left (right) lead is fixed at $T_{\rm L} = 0.41$~K ($T_{\rm R} = 0.01$~K) implying a
  thermal energy $k_B T_{\rm L} = 0.35$~meV ($k_B T_{\rm R} = 0.00086$~meV), respectively.       
  The magnetic field is $B = 10^{-5}$~T, and $\hbar \Omega_0 = 1.0~{\rm meV}$.}
\label{fig07}
\end{figure}

Figure \ref{fig07} indicates the TL-Thermospin current (a) and CL-Thermospin current (b) of the quantum ring system
versus the RSO-coupling for both FES and SES. The photon field causes to decrease the CL-Thermospin current 
while the TL-Thermospin current is almost unchanged. The suppression of CL-Thermospin of the FES, off-resonant regime, 
is due to the enhancement of the spin-polarization shown in \fig{fig06} especially in the ranges 
of the degenerate energy states ($\alpha = [10-15]$~meV nm).

It is interesting to see the resonant regime.  The spin polarization of the SES is decreased in 
the presence of the cavity as is shown in \fig{fig06}, the CL-Thermospin current of the SES is also decreased. 
In addition to the $S_y$ spin polarization that is used to explain the characteristics of the thermospin current, 
the photon replica states play 
an important role in the properties of the thermospin current. In the resonant regime, the $\gamma$GS together with a SES 
form a Rabi-split pair (see \fig{fig02}) and contribute to the transport. 
As a result, the participation of the $\gamma$GS to the transport decreases the CL-Thermospin current.

\begin{figure}[htb]
\centering
    \includegraphics[width=0.45\textwidth,angle=0,bb=50 95 410 210]{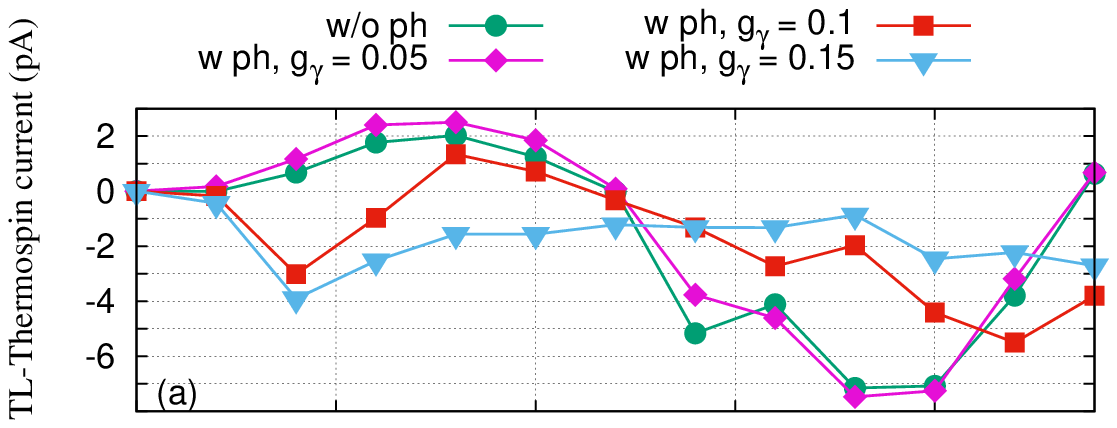}\\
    \includegraphics[width=0.45\textwidth,angle=0,bb=65 55 408 205]{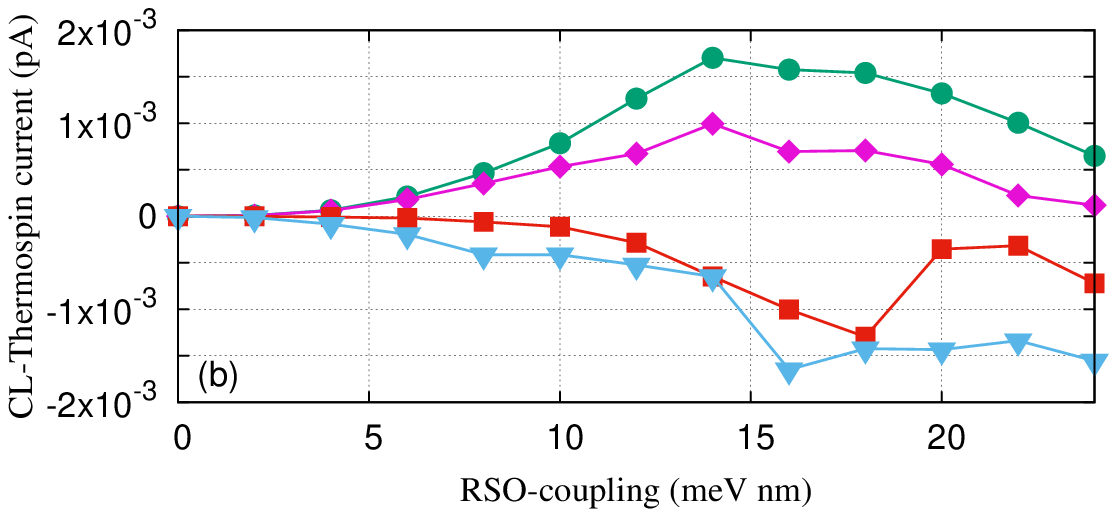}\\
 \caption{(Color online) (a) TL-Thermospin current and (b) CL-Thermospin current versus the RSO-coupling are plotted 
  for the SES (on-resonant regime) of the quantum ring system without photon field (w/o ph) (green circles)
  and with photon field (w ph) for the electron-photon coupling strength $g_{\gamma} = 0.05$ (magenta diamonds),
  $0.1$ (red rectangles), and $0.15$~meV (blue triangles).
  The photon energy is $\hbar \omega_{\rm \gamma} = 0.55$~meV.
  The temperature of the left (right) lead is fixed at $T_{\rm L} = 0.41$~K ($T_{\rm R} = 0.01$~K) implying a
  thermal energy $k_B T_{\rm L} = 0.35$~meV ($k_B T_{\rm R} = 0.00086$~meV), respectively.       
  The magnetic field is $B = 10^{-5}$~T, and $\hbar \Omega_0 = 1.0~{\rm meV}$.}
\label{fig08}
\end{figure}

To further show the influences of the photon field on thermospin transport, we 
present \fig{fig08} which shows the TL-Thermospin current (a) and CL-Thermospin current (b) 
of the SES (on-resonance regime) as a function of the RSO-coupling for different values of the electron-photon coupling strength $g_{\gamma}$. 
Both the TL-Thermospin and CL-Thermospin currents are suppressed with increasing electron-photon coupling strength.
This reduction in the TL- and CL-Thermospin currents is
a direct consequence of the Rabi-splitting of the energy levels
of the quantum ring system in which the energy spacing between the SES and $\gamma$GS shown 
in \fig{fig02} and \fig{fig03}(b) is increased at high electron-photon coupling strength~\cite{Nzar_ACS2016}.
Therefore, the TL- and CL-Thermospin currents decrease.

\section{Conclusions}
\label{Sec:IV}

In summary, we have demonstrated properties of a thermospin transport 
through a quantum ring coupled to a photon cavity.  
In the absence of the photon field, thermospin current is induced at a low temperature gradient 
of the reservoirs that are connected to the quantum ring system. 
Tuning the Rashba spin-orbit coupling, degenerate energy states are formed. 
It is observed that spin-polarization is maximum at the point of degenerate energy states corresponding to 
the AC destructive interference. In the presence of the photon field, the thermospin transport can be controlled using 
a single photon mode in the cavity. Two regimes, off- and on-resonant regimes, are studied.
In the resonant regime, when the photon energy is approximately 
equal to the two lowest energy state of the quantum ring system, photon replica states are formed and 
the spin polarization is sufficiently enhanced.
Tuning the electron-photon coupling strength, the energy spacing between the states is increased 
leading to a suppression of the thermospin transport which is a direct consequence of the Rabi-splitting.

\section{Acknowledgment}
This work was financially supported by the Research
Fund of the University of Iceland, the Icelandic Research
Fund, grant no. 163082-051, and the Icelandic Infrastructure Fund. 
We acknowledge the University of Sulaimani, and Komar Research Center, 
Komar University of Science and Technology, Sulaimani City, Iraq

\bibliographystyle{elsarticle-num} 

\end{document}